\documentclass[conference]{IEEEtran}

\usepackage{hyperref}
\usepackage{amsmath}
\usepackage{multirow}

\newsavebox{\tablebox}
\usepackage{graphicx}
\graphicspath{{image/}}
\usepackage{todonotes} 
\usepackage{wrapfig} 
\usepackage[numbers,sort&compress,square]{natbib} 

\usepackage{rotating} 

\IEEEoverridecommandlockouts 

\begin{document}%

\title{{Automated Behavioral Analysis of Malware}\\ 
\huge{A Case Study of WannaCry Ransomware}\thanks{\footnotesize{This manuscript has been authored by UT-Battelle, LLC under Contract No. DE-AC05-00OR22725 with the U.S. Department of Energy. The United States Government retains and the publisher, by accepting the article for publication, acknowledges that the United States Government retains a non-exclusive, paid-up, irrevocable, world-wide license to publish or reproduce the published form of this manuscript, or allow others to do so, for United States Government purposes. The Department of Energy will provide public access to these results of federally sponsored research in accordance with the DOE Public Access Plan (http://energy.gov/downloads/doe-public-access-plan).
This material is based in part upon work supported by the National Science Foundation under Grant No.1700391. Any opinions, findings, and conclusions or recommendations expressed in this material are those of the authors and do not necessarily
 reflect the views of the National Science Foundation.}}
}

 \author{\IEEEauthorblockN{Qian Chen}
 \IEEEauthorblockA{Department of Electrical and Computer Engineering\\
 The University of Texas at San Antonio\\
 San Antonio, TX 78249\\
 guenevereqian.chen@utsa.edu}
 \and
 \IEEEauthorblockN{Robert A. Bridges}
 \IEEEauthorblockA{Computational Sciences and Engineering Division\\ 
 Oak Ridge National Laboratory\\
 Oak Ridge, TN 37831\\
 bridgesra@ornl.gov}
 }


\maketitle

\begin{abstract}
Ransomware, a class of self-propagating malware that uses encryption to hold the victims' data ransom, has emerged in recent years as one of the most dangerous cyber threats, with widespread damage;  
e.g., zero-day ransomware WannaCry has caused world-wide catastrophe, from knocking U.K. National Health Service hospitals offline to shutting down a Honda Motor Company in Japan~\cite{Honda2017}.  
Our close collaboration with security operations of large enterprises reveals that defense against ransomware relies on tedious analysis from high-volume systems logs of the first few infections. 
Sandbox analysis of freshly captured malware is also commonplace in operation. 

We introduce a method to identify and rank the most discriminating ransomware features from a set of ambient (non-attack) system logs and at least one log stream containing both ambient and ransomware behavior. 
These ranked features reveal a set of malware actions that are produced automatically from system logs, and can help automate tedious manual analysis. 
We test our approach using WannaCry and two polymorphic samples by producing logs with Cuckoo Sandbox during both ambient, and ambient plus ransomware executions. 
Our goal is to extract the features of the malware from the logs with only knowledge that malware was present. 
We compare outputs with a detailed analysis of WannaCry allowing validation of the algorithm's feature extraction and provide analysis of the method's robustness to variations of input data\textemdash changing quality/quantity of ambient data and testing polymorphic ransomware. 
Most notably, our patterns are accurate and unwavering when generated from polymorphic WannaCry copies, on which 63 (of 63 tested) anti-virus (AV) products fail. 
\end{abstract}


\section{Introduction}
\label{sec:intro}
Ransomware is a class of self-propagating malware that uses encryption to hold victim's data ransom and has emerged as a dominant worldwide threat, crippling personal, industrial, and governmental networked resources~\cite{cryptowall, CryptoRansom, Murnane2017The, Malwarebytes2017}. 
Most notably, the recent epidemic of WannaCry~\cite{CERT2017WannaCry, WannaCryHonda} was one of the largest ransomware attacks in history,  halting hospital facilities and infecting large corporations and consumers in over 150 countries. 
From initial exploit to completing encryption of a host's data, ransomware must perform a series of actions; e.g., identifying files for encryption/deletion and exchanging encryption keys with a command and control (CC) server. 
Hence, discovery of the malware's pre-encryption footprint promises accurate, in-time detection and is the focus of ransomware analysis efforts. 
Our hypothesis is that data analytics on host logs can automate discovery of features that are indicative of ransomware's presence before encryption (and of malware's executions in general). 
Such a capability promises automated pattern-generation and analysis capabilities that are robust to syntactic polymorphism in ransomware and more general classes of malware; a critical necessity given the unfortunate success of the ransomware economy.  

This work is motivated by our close collaboration with cyber operations at large enterprises.  
Most notably the 2015 infection by polymorphic and then novel CryptoWall 3.0 induced a 160 man-hour forensic effort to manually analyze the few infected hosts'  logs in an attempt to produce shareable threat intelligence reports and pre-encryption detection capabilities. 
This operational scramble begs the question, ``How to automatically extract the sequence of events induced by malware given a large volume of logs from a few hosts that were infected with potentially polymorphic malware?''
Moreover, such facilities regularly receive freshly discovered malware samples, which are analyzed in sandboxes; 
hence, an automated pattern-generation tool that is provably (more) robust to polymorphism from dynamic analysis is needed.

To our knowledge no automated method for extracting the footprint of malware from the ambient and/or sandbox-generated logging data is known. Yet our close collaboration with security analysts reveals that such a method is needed in practice to benefit two primary use cases\textemdash (1) to expedite currently timely (hundreds of man hours) manual analysis of logs used to identify malware’s actions from ambient system logs in forensic efforts, (2) to automatically generate behavioral analysis of malware samples from sandbox logging data (which is currently investigated manually). Our contribution is to present an algorithm for automatically extracting the features that discriminate malware from ambient logging activity given collections of logs known to contain malware executions (in addition to ambient logs) and logging data known to contain no malware executions. Further, we present systematic testing of our method using Cuckoo Sandbox to generate WannaCry ransomware logs showing when and how it is robust to changes in input data. 

To this end, we propose an information relative approach using a Term Frequency-Inverse Document Frequency (TF-IDF) metric to automatically extract and rank the most discriminating features of the new malware from logging data. The TF-IDF method also preserves human-understandable features, which is necessary for operators to understand their analytics, e.g., in  automatic malware analysis.

We leverage Cuckoo Sandbox (\url{https://cuckoosandbox.org}), 
an automatic malware analysis system, for dynamic analysis of executables including both WannaCry variants and scripts simulating non-malicious user activity.  
Cuckoo reports activity relating to files,  folders, memory, network traffic, processes, and API calls, thus giving source data for experimentation with ground truth. 
This builds results on a well-adopted, open-source malware analysis tool (Cuckoo), ensures repeatability of results, and proves a concept that we believe will be transferable to more general system logs, e.g., Windows Logging Service output (\url{https://digirati82.com/wls-information}) that are not shareable outside the organization. 

The algorithm's outputs are validated using a detailed analysis of WannaCry. 
Our contributions include feature extraction techniques from Cuckoo output, and, most notably, a method to automatically extract the most discriminative ransomware features from event logs given a set of ambient  (non-attack) logs and logs containing ransomware (potentially mixed with logs from many normal activities).  
We present four experiments showing our method (1) can automatically extract features that are indicative of the malware, (2) the method is robust to the quantity of known, non-malicious logging data included, (3) the method succeeds when the logs containing malware activity also include a majority of non-malicious ambient logs, and (4) our method can produce a pattern that is robust to polymorphic changes that bypass 63 (of 63 tested) anti-virus (AV) detectors.



\subsection{Related Work} 
\label{sec:related}
Malware analysts usually adopt static and dynamic analysis techniques to determine behavior and risks of a specific malware sample. 
Static analysis analyzes malware samples before it is executed~\cite{awad2016automatic} but struggles to analyze self-modifying and polymorphic code. 
Dynamic analysis is more powerful for malware forensics analysis because it allows analysts to understand malware behavior and activities by executing the malware sample. In this work, we use Cuckoo Sandbox for dynamic analysis. 

Cuckoo has been used to identify polymorphic malware samples~\cite{provataki2013differential}, trigger malware that detects it is in a sandbox, and identifies particular malware actions in different network profiles~\cite{Ceron2016mars}, find IP addresses, domains and file hashes of malware samples to generate network-related indicators of compromise~\cite{Rudman2016Dridex}, and providing ground-truth training and testing data for supervised Intrusion Detection System (IDS) approaches~\cite{Shijo2014Integrated, Lim2014Mal-ONE, Vasilescu2014Practical, Mosli2016Automated, Kawaguchi2015Malware, Kumar2016Machine, Qiao2013Analyzing, Hansen2016An, Pektaş2015runtime, Kirat2014BareCloud, Fujino2015Discovering}. Therefore, Cuckoo is an established tool to generate repeatable  malware analysis results.

UNVEIL~\cite{kharraz2016unveil} is a novel dynamic analysis system for detecting ransomware attacks and modeling their behaviors. UNVEIL tracks changes to the analysis system’s desktop by calculating dissimilarity scores of desktop screenshots before, during, and after executing the malware samples to identify ransomware.  UNVEIL successfully identified previously unknown evasive ransomware that was not detected by the anti-malware software. 

UNVEIL focuses on ransomware attack detection while our work is a generic approach to extract the most discriminative features of malware in logging data. 
We specially tested our approach with WannaCry malware. Our approach automatically discovers features that are indicative of WannaCry ransomware's presence before encrypting targeted files.
\begin{wraptable}{r}{.28\textwidth}
    \vspace{-.2cm}
    \centering
    \caption{WannaCry Files \& Folders}
    \label{tab:files}
    \begin{lrbox}{\tablebox}{}
    \begin{tabular}{|c|l|l|}
    \hline
    \textbf{Name} & \textbf{Meaning} \\
        b.wnry &  Bitmap file for Desktop image\\ \hline
        c.wnry &  Configuration file \\ \hline
        r.wnry &  Q\&A file, payment instructions \\ \hline
        s.wnry &  Tor client\\ \hline
        t.wrny &  WANACRY! file with RSA keys \\ \hline
        u.wnry &  @WannaDecryptor@.exe \\ \hline
        \multirow{3}{*}{\textbackslash msg}  
            & Folder containing RTF files with\\
            & payment instructions in 128\\
            & languages  (e.g., korean.wnry)\\ \hline
        taskse.exe & Launches decryption tool\\ \hline
        taskdl.exe & Removes temporary files\\  
    \hline
    \end{tabular}
    \end{lrbox}
\scalebox{0.7}{\usebox{\tablebox}}
\vspace{-.9cm}
\end{wraptable}
\section{Prerequisites} 
\subsection{WannaCry Ransomware Attack}
\label{sec:malware}
Our primary goal is to extract malware's activity from a set of logs only knowing the logs contain malware activity and thereby automating malware analysis and pattern generation. 
Before testing the capability, we present an overview of WannaCry to be used as ground truth for validating the malware features extracted from our tests. 
In May 2017, the WannaCry ransomware attack infected over 300k Windows computers in over 150 countries. 
\begin{wrapfigure}{r}{.3\textwidth}
    \vspace{-.2cm}
    \centering
    \includegraphics[scale=0.4]{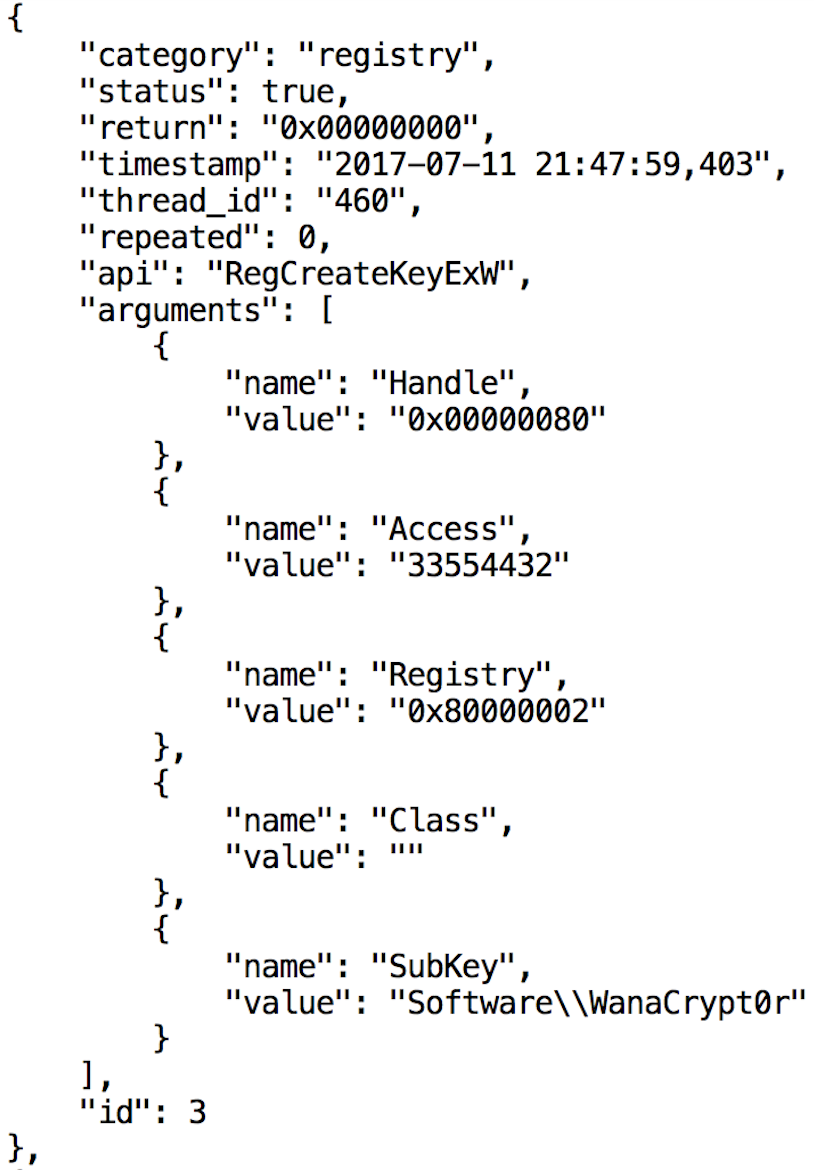}
    \caption{Behavioral Log Example}
    \label{fig:registrylog}
    \vspace{-.2cm}
\end{wrapfigure}
The dropper of the malware carries two components. 
One uses the ``EternalBlue'' exploit against a vulnerability of Windows' Server Message Block (SMB) protocol to propagate, and the other is a WannaCry ransomware encryption component~\cite{CERT2017WannaCry}. Static analysis of WannaCry has been documented by analysts and cyber security companies~\cite{WannaCry2017}. 
The analyzed WannaCry’s files and action sequence are summarized in Tables~\ref{tab:files} and \ref{tab:actions}, respectively. 

More details are available in \hyperref[sec:appendix-wannacry]{Appendix A}.
This gives ground-truth for evaluating the features identified from Cuckoo logs.

\subsection{Cuckoo Sandbox \& Produced Logs}
\label{subsec:Cuckoo}
Cuckoo Sandbox is an automatic malware analysis system, which provides detailed results of suspicious files' activities and behaviors by executing the files (e.g., Windows executables, document exploits, URLs and HTML files, Java JAR, ZIP file, Python files etc.) in a virtualized and isolated environment. 

The suspicious file's performance, such as changes of files and folders, memory dumps, network traffic, processes and the API calls are monitored and analyzed by Cuckoo. 
The Cuckoo reporting module elaborates the analysis results and saves the produced report into a human readable JavaScript Object Notation (JSON) and HTML formats.
In our experiments Cuckoo outputs ranged from 1MB to 1GB per analyzed file.


\begin{table}[!ht]
\vspace{-.2cm}
\centering
\caption{WannaCry Actions}
\label{tab:actions} 
\begin{lrbox}{\tablebox}{}
\begin{tabular}{|c|c|l|}
\hline
\multirow{7}{*}{\begin{sideways}Pre-Encryption  \phantom{aaaaaaaaa} \end{sideways}} & \textbf{No.} & \textbf{Action} \\ \cline{2-3} 
 & 1 & Imports CryptoAPI from advpi32.dll \\ \cline{2-3} 
 & 2 & Unzips itself to files in Table~\ref{tab:files}\\ \cline{2-3} 
 & 3 & Generates machine-unique identifier\\ \cline{2-3} 
 & 4 & \begin{tabular}[c]{@{}l@{}} 
        Creates a registry, \path{HKEY_LOCAL_MACHINE\}\\ \path{Software\WanaCrypt0r\wd}
        \end{tabular} \\ \cline{2-3}  
 & 5 &  \begin{tabular}[c]{@{}l@{}}  
        Runs `attrib +h', which sets the\\
        current directory as a hidden folder
        \end{tabular} \\ \cline{2-3} 
 & 6 &  \begin{tabular}[c]{@{}l@{}}
            Runs `icacls . /grant Everyone:F /T /C \\
            /Q', which grants all users permissions \\
            to the current directory
        \end{tabular} \\ \cline{2-3} 
 & 7 &   \begin{tabular}[c]{@{}l@{}}
            Imports public and private RSA AES \\
            keys (000.pky, 000.eky) from t.wrny\\ 
        \end{tabular} \\ \cline{2-3} \hline
 \multirow{5}{*}{\begin{sideways} Encryption  \phantom{aaaaaaaaaaa} \end{sideways}} 
 & 8 & \begin{tabular}[c]{@{}l@{}} 
        Creates  00000000.res, a file containing unique user ID,
        \\total encrypted file count,
        and total encrypted file size
        \end{tabular} \\ \cline{2-3} 
 & 9 & \begin{tabular}[c]{@{}l@{}} 
        Uses SHGetFolderPathW API to scan the file system \\
        starting at the desktop folder
        \end{tabular} \\ \cline{2-3}
 & 10 & \begin{tabular}[c]{@{}l@{}} 
        Finds the target files and generates one AES key per file
        \end{tabular} \\ \cline{2-3}
        
 & 11 & \begin{tabular}[c]{@{}l@{}} 
        Uses the public RSA key to encrypt AES key of target\\ file and saves encrypted AES key to the target file 
        \end{tabular} \\ \cline{2-3}
        
& 12 & \begin{tabular}[c]{@{}l@{}} 
        Uses CreateFileW, ReadFile and WriteFile APIs to create\\ encrypted files. The string “WANNACRY” is written on\\
        the infected files.
        \end{tabular} \\ \cline{2-3}
        
 & 13 & \begin{tabular}[c]{@{}l@{}} 
        Calls taskdl.exe and MoveFileW API to replace\\
        .WNCRYT temp files to .WCRY files
        \end{tabular} \\ \cline{2-3}
        
 & 14 & \begin{tabular}[c]{@{}l@{}}
        Calls taskse.exe \path{C:\DOCUME~1\cuckoo}\\ \path{\LOCALS~1\Temp\@WanaDecryptor@.exe}\\
        to launch the decryption tool and\\
        replace desktop image to ``!WannaCryptor!.bmp"
        \end{tabular}\\ \cline{2-3}  
 & 15 & \begin{tabular}[c]{@{}l@{}}
        Runs \path{cmd.exe /c reg add HKLM\SOFTWARE\}\\ \path{Microsoft\Windows\CurrentVersion\Run}\\ \path{ /v \"thsgvkvtwaipdcd971\" /t REG_SZ /d\}\\ \path{"\"C:\DOCUME~1\cuckoo\LOCALS~1\}\\ \path{Temp\tasksche.exe\\"\" /f} to create \\
        a unique identifier registry key
        \end{tabular}\\ \cline{2-3}
 & 16 & 
    Runs @WanaDecryptor@.txt  to create a copy of
    r.wnry 
    \\ \cline{2-3} 
 & 17 & Temporary files with prefix `$\sim$SD' created then deleted \\ \hline
\end{tabular}
\end{lrbox}
\scalebox{0.8}{\usebox{\tablebox}}
\vspace{-.35cm}
\end{table}

Although Cuckoo reports seven categories of malware analysis outputs,  we restrict ourselves to the \textit{behavior} category, as these are analogous to system logs collected by security operations from workstations.
This category contains the raw behavioral logs for each process running by the analyzed files, including logs of the complete processes tracing, a behavioral summary and a process tree.  

The \textit{enhanced} class generates a more extensive high-level summary of the processes and their activities.Instead of reading from raw behavioral logs, the enhanced class helps to interpret and summarize essential activities performed by the analyzed files, e.g.,  
read, write, and delete from registry, files, and directories; load Windows libraries; and execute files. 

\begin{figure}[!ht]
\vspace{-.3cm}
\includegraphics[scale=0.34]{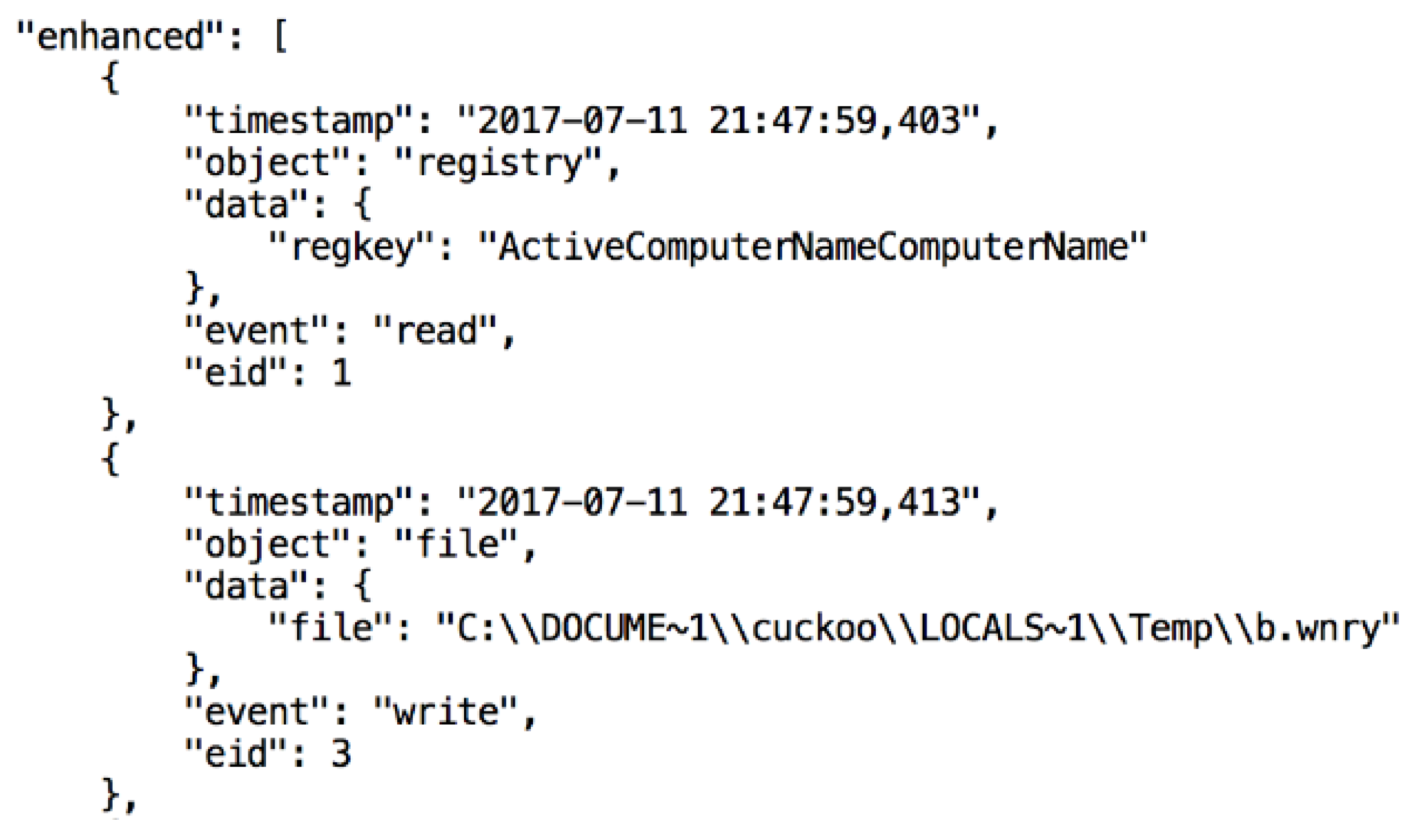}
\centering
\caption{Enhanced logs of a Cuckoo Analysis JSON Report File}
\label{fig:enhanced}
\vspace{-.4cm}
\end{figure}

See Fig.~\ref{fig:enhanced}. 
Additionally, Cuckoo searches \url{VirusTotal.com}, checking 63 AV vendors for signatures detecting the file under analysis. 
We leverage this  to compare the capabilities of our pattern generation for polymorphic samples to that of AV vendors. 
See \hyperref[sec:appendix-cuckoo]{Appendix B} for more information on Cuckoo.

\begin{table}[]
    \centering
    \caption{Extracted Feature Format for Logs in Figures \ref{fig:registrylog} \& \ref{fig:enhanced}}
    \label{tab:FeatureExtracted}
    \begin{lrbox}{\tablebox}
    \begin{tabular}{|c|l|l|}
    \hline
    \multicolumn{1}{|l|}{Log Examples} & Feature                                                                                                                                                                                                     \\ \hline
     
    Fig.~\ref{fig:registrylog}   & ``bigram:\_api=regcreatekeyexw+arguments=software\path{\\}wanacrypt0r"\\
    \hline
    
    \begin{tabular}[c]{@{}l@{}}
    Fig.~\ref{fig:enhanced}\\(``eid":1)
    \end{tabular}
    & \begin{tabular}[c]{@{}l@{}}``enhanced:\_object=registry+event=read+data=\\regkey:activecomputernamecomputername "\end{tabular}
    \\ \hline

     \begin{tabular}[c]{@{}l@{}}
     Fig.~\ref{fig:enhanced}\\(``eid":3)  
     \end{tabular}
     & \begin{tabular}[c]{@{}l@{}}``enhanced:\_object=file+event=write+data=\\
    file:c:$\setminus\setminus$docume$\sim$1$\setminus\setminus$cuckoo$\setminus\setminus$locals$\sim$1$\setminus\setminus$temp$\setminus\setminus$b.wnry "\end{tabular}
    \\ \hline
    \end{tabular}
    \end{lrbox}
    \scalebox{0.85}{\usebox{\tablebox}}
    \vspace{-.2cm}
\end{table}


\section{Method: Features \& TF-IDF} 
\label{sec:method} 
The general problem we consider is how to extract the most indicative features of malware from logs of the host  on which the malware was active. 
Note that this set of logs may contain a majority of logs from non-malicious, ambient user activity. 
Our approach is to obtain a second set of logs from only non-malicious activity (e.g., by creation in our case, or in the operational case of a ransomware outbreak, from system logs of non-infected hosts), and seek features of the infected logs set that are uncommonly common (are high frequency in a few malicious documents only). 
Below we describe the feature representation from Cuckoo behavior logs and the ranking method. 

Conceptually, we consider a set of logs (Cuckoo enhanced and behavior logs in our experiments) as a ``document'' and a selected subset of the log entries as ``terms'' or features. 
Two entries are considered the same term/feature if they agree in all fields except time and event ID. 
All enhanced logs are used. 
Only those behavior (non-enhanced) logs with the fields ``category'' and ``api'' taking values \textit{registry} and \textit{RegCreatKeyExW}, respectively, are included. 
Altogether, a document (log stream) is represented as a count of each term/feature (a bag-of-words model). 
See Table~\ref{tab:FeatureExtracted}. 



For application to ransomware pattern generation, we only consider pre-encryption features; specifically, for WannaCry logs 
before the creation of the private key 00000000.eky. 
In practice, given the initial infection host logs, operators would have to identify a pre-infection cutoff and apply our method to all logs previous. 
Although for general malware forensics / analysis this is not necessary. 
Given two sets of documents (in our case, at least one with logs containing malware activities, the some containing ambient activity), we apply Term-Frequency-Inverse-Document-Frequency (TF-IDF) an information relative term-weighting scheme~\cite{Salton1988Term}. 
Letting $f(t,d)$ denote the frequency of term $t$ in document $d$, and $N$ the size of the corpus, the TF-IDF weight is the product of the Term Frequency,  $\mbox{tf}(t,d)=f_{t,d}/\sum_{t'\in d} f_{t',d}$ 
(giving the likelihood of $t$ in $d$) and the Inverse Document Frequency,
$\mbox{idf}(t,D) = \log {N}/{1+|\{d\in D: t\in d\}|}$ (giving the Shannon's information of the a document containing  $t$). 
Intuitively, given a document, those terms that are uncommonly high frequency in that document are the only that receive high scores. 

Our application is to consider all logs from infected hosts as a single document, then regard only the features from this ``infected'' document and apply TF-IDF; hence, highly ranked features occur often in  (and are guaranteed to occur at least once in) the ``infected''  document, but infrequently anywhere else.

\section{Experiments \& Results}
\label{sec:experiment}

\subsection{Analysis of WannaCry \& Four Normal Activities}
\label{subsec:experiment1}
In our first experiment, we first analyze malicious behavior of the WannaCry executable file by sending it to the Cuckoo Sandbox. 
Besides obtaining a Cuckoo analysis report of the WannaCry sample (i.e., a malicious document), Python scripts of users' normal activity (i.e. read, write and delete files, open websites, watch YouTube videos, send and receive emails, search flight tickets, post and delete tweets on Twitter) are submitted to and executed by Cuckoo. One WannaCry malicious document and four normal documents (Cuckoo analysis reports of users' normal activity) are used to calculate TF-IDF weights for 74 pre-encryption WannaCry specific features.  
Note that the normal activity analysis reports contain various features that may or may not be in the malware analysis report,
and some of the 74 features extracted from the malicious logs may have occurred from ambient (non-malicious) behavior (see Section~\ref{subsec:experiment3} where this is known).  

Table~\ref{weight1} shows the most important 43 features  (top-ten TF-IDF weights). These highly ranked features are also the patterns of WannaCry obtained from the detailed technical analysis of the WannaCry executable file (Section~\ref{sec:malware}).  

\begin{table}
\vspace{-.3cm}
\centering
\caption{The Ranking of Features and Their TF-IDF Weights}
\label{weight1}
\begin{lrbox}{\tablebox}
\begin{tabular}{|c|l|c|}
\hline
Ranking                 & Feature                                                                                                                                            & \begin{tabular}[c]{@{}l@{}}TF-IDF\\Weight\end{tabular}                \\ \hline
\multicolumn{1}{|c|}{1} & \begin{tabular}[c]{@{}l@{}}``enhanced:\_object=file+event=write+data=file:\\c:\path{\\}docume$\sim$1\path{\\}cuckoo\path{\\}locals$\sim$1\path{\\}temp\path{\\}s.wnry"\end{tabular}  & \multicolumn{1}{c|}{299.36} \\ \hline
\multicolumn{1}{|c|}{2} & \begin{tabular}[c]{@{}l@{}}``enhanced:\_object=file+event=write+data=file:\\c:\path{\\}docume$\sim$1\path{\\}cuckoo\path{\\}locals$\sim$1\path{\\}temp\path{\\}b.wnry"\end{tabular}  & \multicolumn{1}{c|}{33.80}  \\ \hline
\multicolumn{1}{|c|}{3} & \begin{tabular}[c]{@{}l@{}}``enhanced:\_object=file+event=write+data=file:\\c:\path{\\}docume$\sim$1\path{\\}cuckoo\path{\\}locals$\sim$1\path{\\}temp\path{\\}u.wnry "\end{tabular} & \multicolumn{1}{c|}{24.14}  \\ \hline
\multicolumn{1}{|c|}{4} & \begin{tabular}[c]{@{}l@{}}``enhanced:\_object=file+event=read+data=file:\\c:\path{\\}docume$\sim$1\path{\\}cuckoo\path{\\}locals$\sim$1\path{\\}temp\path{\\}t.wnry "\end{tabular}  & \multicolumn{1}{c|}{9.66}   \\ \hline
5       & \begin{tabular}[c]{@{}l@{}}``enhanced:\_object=file+event=write+data=file:\\c:\path{\\}docume$\sim$1\path{\\}cuckoo\path{\\}locals$\sim$1\path{\\}temp\path{\\}msg\path{\\}m\_korean.wnry ",\\``enhanced:\_object=file+event=write+data=file:\\
c:\path{\\}docume$\sim$1\path{\\}cuckoo\path{\\}locals$\sim$1\path{\\}temp\path{\\}msg\path{\\}m\_vietnamese.wnry" \end{tabular} & 9.66                      \\ \hline

6      & \begin{tabular}[c]{@{}l@{}}``enhanced:\_object=file+event=write+data=file:\\ c:\path{\\}docume$\sim$1\path{\\}cuckoo\path{\\}locals$\sim$1\path{\\}temp\path{\\}msg\path{\\}m\_chinese (traditional).wnry", \\
``enhanced:\_object=file+event=write+data=file:\\ c:\path{\\}docume$\sim$1\path{\\}cuckoo\path{\\}locals$\sim$1\path{\\}temp\path{\\}msg\path{\\}m\_japanese.wnry" \end{tabular}                                                                                                                                                                                                                                                                               & 8.05                      \\ \hline

7      & \begin{tabular}[c]{@{}l@{}}"enhanced:\_object=file+event=write+data=file:\\ c:\path{\\}docume$\sim$1\path{\\}cuckoo\path{\\}locals$\sim$1\path{\\}temp\path{\\}msg\path{\\}m\_chinese (simplified).wnry",\\
\ldots (\textbf{24 various language features}) \\ 
``enhanced:\_object=file+event=write+data=file:\\ c:\path{\\}docume$\sim$1\path{\\}cuckoo\path{\\}locals$\sim$1\path{\\}temp\path{\\}msg\path{\\}m\_turkish.wnry" \end{tabular}                                                                                                                                                                                                                                                                                           & 4.83                      \\ \hline

8      & \begin{tabular}[c]{@{}l@{}}``enhanced:\_object=file+event=read+data=file:\\ c:\path{\\}docume$\sim$1\path{\\}cuckoo\path{\\}locals$\sim$1\path{\\}temp\path{\\}c.wnry ",\\
``enhanced:\_object=file+event=write+data=file:\\ c:\path{\\}docume$\sim$1\path{\\}cuckoo\path{\\}locals$\sim$1\path{\\}temp\path{\\}c.wnry",\\
``enhanced:\_object=file+event=write+data=file:\\ c:\path{\\}docume$\sim$1\path{\\}cuckoo\path{\\}locals$\sim$1\path{\\}temp\path{\\}taskdl.exe",\\
``enhanced:\_object=file+event=write+data=file:\\ c:\path{\\}docume$\sim$1\path{\\}cuckoo\path{\\}locals$\sim$1\path{\\}temp\path{\\}taskdl.exe",\\
``enhanced:\_object=registry+event=read+data=regkey:\\
\path{\\} activecomputernamemachineguid"\end{tabular}              & 3.22                      \\ \hline

9      & \begin{tabular}[c]{@{}l@{}}``enhanced:\_object=registry+event=read+data=regkey:\\
hkey\_local\_machine\path{\\}software\path{\\}microsoft\path{\\}cryptography\path{\\}defaults\path{\\}provider\path{\\}  \\
microsoft enhanced rsa and aes cryptographic provider (prototype)image path"\end{tabular}                                                                                                                                                                                                                                                                                                                               & 2.04                      \\ \hline

10      & \begin{tabular}[c]{@{}l@{}}``bigram:\_api=regcreatekeyexw+arguments=software\path{\\}wanacrypt0r",\\
``enhanced:\_object=dir+event=create+data=file:\\ 
c:\path{\\}docume$\sim$1\path{\\}cuckoo\path{\\}locals$\sim$1\path{\\}temp\path{\\}msg",\\
``enhanced:\_object=file+event=execute+data=file:attrib +h . ",\\
``enhanced:\_object=file+event=execute+data=file:icacls . /grant everyone:f /t /c /q ",\\ ``enhanced:\_object=file+event=write+data=file:\\
c:\path{\\}docume$\sim$1\path{\\}cuckoo\path{\\}locals$\sim$1\path{\\}temp\path{\\}00000000.pky",\\
``enhanced:\_object=file+event=write+data=file:\\
c:\path{\\}docume$\sim$1\path{\\}cuckoo\path{\\}locals$\sim$1\path{\\}temp\path{\\}r.wnry" \end{tabular} & 1.61                      \\ \hline

\end{tabular}
\end{lrbox}
\scalebox{0.65}{\usebox{\tablebox}}
\vspace{-.5cm}
\end{table}

\subsection{Analysis of WannaCry \&  Varying Normal Activities}
\label{subsec:experiment2}
This experiment aims to validate that the ranking of WannaCry features is not influenced by varying the number of normal documents. 
To validate the hypothesis, we calculate the TF-IDF weights in the following three scenarios.

\begin{enumerate}
\item One analysis report of the WannaCry executable file and five normal performance analysis files.
\item The same analysis report of the WannaCry executable file and six normal performance analysis files.
\item The same analysis report of the WannaCry executable file and 17 normal performance analysis files.
\end{enumerate} 
Normal activities are analyzed by Cuckoo Sandbox via submitting Python files.   From the three experiments we find that the highest 10 weights calculated by TF-IDF and their features \textit{are the same as} the ranking shown in Table~\ref{weight1}, regardless of the number of normal activity analysis files.

\subsection{Combining Normal Activities with WannaCry} 
\label{subsec:experiment3}
In this experiment, we create a Python file that executes normal activities first and then trigger the WannaCry malware. The Python file is analyzed by the Cuckoo Sandbox, and the analysis report along with various non-malicious log files are sent to the TF-IDF method. 
This experiment aims to \textit{validate that our method can accurately identify specific features of the malware when a large majority of the features are indicative of non-malicious activity}. 
In operations, this gives evidence that our method can help IT staff distinguish the malware's footprint from  majority ambient logging data. 

To create the mixed normal + malware logs, we use a Python script to search flight ticket information by opening the website (i.e., \path{www.google.com/flights}) and changing the date and airport codes of the URL link for various requests. After the normal activities, the system executes the WannaCry executable files. The Python script combining both normal and WannaCry activities is submitted to the Cuckoo analyzer, and the report of the analysis is used to calculate and determine the most important patterns of the combination scenario. 

We still search for the timeline when the private key was first generated, and consider all features before the timeline (pre-encryption features only). 
Since the Python file executes normal activities (i.e., search for flight information)  before the WannaCry malware, the number of pre-encryption features has increased from 74 to 1,085. 

Two experiments are designed as follows. 
\begin{enumerate}
\item One combination analysis of flight-search (normal) activities and WannaCry malware versus four flight-search normal activity analysis reports.
\item The same combination analysis report versus 21 normal activities (including four normal analysis reports used in the above experiments).
\end{enumerate}

As we introduced in Section~\ref{sec:method}, the ``IDF'' term down weights the features occurring in many documents; hence, although most of the 1,085 features appearing in the normal+malware reports are not indicative of WannaCry, 
the malware-specific features are still ranked as top features, but the ranking of some malware features are scaled down. 
This is because some normal activities appear frequently in the combination analysis report, but relatively infrequently in other normal reports, 
especially in Scenario Two where many different normal activities are included. 
The ranking of the features for two experiments conducted in this Experiment are shown in Table~\ref{weight2}. 
We only list the rankings of the malware specific features in Table~\ref{weight1}.  

For example,  \path{"enhanced:_object=file+event=write+data=file:c:\\documents and settings\\cuckoo\\application data\\mozilla\\firefox\\profiles\\qk4ev1cw.default\\places.sqlite"}, is a Firefox activity for reading and writing the  ``places.sqlite" file to save browsing history, store bookmarks, annotations etc. 
As it is a common activity in all analysis reports where Firefox applications are executed, it is frequent in the combination analysis file (increasing TF), but occurs in (only) 14 out of 21 normal performance analysis (so IDF is not too small). 
Therefore, in the second scenario, the TF-IDF weight of the feature is $69.1$, which is higher than many of the WannaCry specific features.

It is easy to mathematically prove, and we have empirically verified that this method will produce false positives if and only if non-malicious features occurring often in the document containing malware  are infrequent elsewhere. 


\begin{table}[]
\centering
\caption{Features Rankings for Experiment~\ref{subsec:experiment3}}
\label{weight2}
\begin{lrbox}{\tablebox}
\begin{tabular}{|l|c|c|}
\hline
Feature & \begin{tabular}[c]{@{}l@{}}Ranking\\(case 1)\end{tabular}  &
 \begin{tabular}[c]{@{}l@{}}Ranking\\(case 2)\end{tabular} \\ \hline
                                                                                                                                                       
\begin{tabular}[c]{@{}l@{}}``enhanced:\_object=file+event=write+data=file:\\c:\path{\\}docume$\sim$1\path{\\}cuckoo\path{\\}locals$\sim$1\path{\\}temp\path{\\}s.wnry"\end{tabular} &1 & 2  \\ \hline 

\begin{tabular}[c]{@{}l@{}}``enhanced:\_object=file+event=write+data=file:\\c:\path{\\}docume$\sim$1\path{\\}cuckoo\path{\\}locals$\sim$1\path{\\}temp\path{\\}b.wnry"\end{tabular}  & 2 &7 \\ \hline 

\begin{tabular}[c]{@{}l@{}}``enhanced:\_object=file+event=write+data=file:\\c:\path{\\}docume$\sim$1\path{\\}cuckoo\path{\\}locals$\sim$1\path{\\}temp\path{\\}u.wnry "\end{tabular} & 4 & 13 \\ \hline

\begin{tabular}[c]{@{}l@{}}``enhanced:\_object=file+event=read+data=file:\\c:\path{\\}docume$\sim$1\path{\\}cuckoo\path{\\}locals$\sim$1\path{\\}temp\path{\\}t.wnry ",\\

``enhanced:\_object=file+event=write+data=file:\\c:\path{\\}docume$\sim$1\path{\\}cuckoo\path{\\}locals$\sim$1\path{\\}temp\path{\\}msg\path{\\}m\_korean.wnry ",\\

``enhanced:\_object=file+event=write+data=file:\\
c:\path{\\}docume$\sim$1\path{\\}cuckoo\path{\\}locals$\sim$1\path{\\}temp\path{\\}msg\path{\\}m\_vietnamese.wnry "\end{tabular}  &7 & 32    \\ \hline 

\begin{tabular}[c]{@{}l@{}}``enhanced:\_object=file+event=write+data=file:\\ c:\path{\\}docume$\sim$1\path{\\}cuckoo\path{\\}locals$\sim$1\path{\\}temp\path{\\}msg\path{\\}m\_chinese (traditional).wnry", \\
``enhanced:\_object=file+event=write+data=file:\\ c:\path{\\}docume$\sim$1\path{\\}cuckoo\path{\\}locals$\sim$1\path{\\}temp\path{\\}msg\path{\\}m\_japanese.wnry" \end{tabular}                                                                                                                                                                                                                                                                               & 8 & 36                   \\ \hline 

\begin{tabular}[c]{@{}l@{}}``enhanced:\_object=file+event=write+data=file:\\ c:\path{\\}docume$\sim$1\path{\\}cuckoo\path{\\}locals$\sim$1\path{\\}temp\path{\\}msg\path{\\}m\_chinese (simplified).wnry",\\
``enhanced:\_object=file+event=write+data=file:\\
c:\path{\\}docume~1\path{\\}cuckoo\path{\\}locals~1\path{\\}temp\path{\\}msg\path{\\}m\_romanian.wnry "\end{tabular}                                                                                                                                                                                                                                                                               &9&40
 \\ \hline 

\begin{tabular}[c]{@{}l@{}} ``enhanced:\_object=file+event=write+data=file:\\
c:\path{\\}docume~1\path{\\}cuckoo\path{\\}locals~1\path{\\}temp\path{\\}msg\path{\\}m\_bulgarian.wnry "\\
\ldots (\textbf{22 various language features}) \\ 
``enhanced:\_object=file+event=write+data=file:\\ c:\path{\\}docume$\sim$1\path{\\}cuckoo\path{\\}locals$\sim$1\path{\\}temp\path{\\}msg\path{\\}m\_turkish.wnry" \end{tabular}                                                                                                                                                                                                                                                                                          &10 & 45                    \\ \hline 

\begin{tabular}[c]{@{}l@{}}``enhanced:\_object=file+event=read+data=file:\\ c:\path{\\}docume$\sim$1\path{\\}cuckoo\path{\\}locals$\sim$1\path{\\}temp\path{\\}c.wnry ",\\
``enhanced:\_object=file+event=write+data=file:\\ c:\path{\\}docume$\sim$1\path{\\}cuckoo\path{\\}locals$\sim$1\path{\\}temp\path{\\}c.wnry",\\
``enhanced:\_object=file+event=write+data=file:\\ c:\path{\\}docume$\sim$1\path{\\}cuckoo\path{\\}locals$\sim$1\path{\\}temp\path{\\}taskdl.exe",\\
``enhanced:\_object=file+event=write+data=file:\\ c:\path{\\}docume$\sim$1\path{\\}cuckoo\path{\\}locals$\sim$1\path{\\}temp\path{\\}taskdl.exe",\\
``enhanced:\_object=registry+event=read+data=regkey:\\
\path{\\} activecomputernamemachineguid"\end{tabular}  
& 12      & 54                   \\ \hline 

\begin{tabular}[c]{@{}l@{}}``enhanced:\_object=registry+event=read+data=regkey:\\
hkey\_local\_machine\path{\\}software\path{\\}microsoft\path{\\}cryptography\path{\\}defaults\path{\\}provider\path{\\}  \\
microsoft enhanced rsa and aes cryptographic provider (prototype)image path"\end{tabular}                                                                                                                                                                                                                                                                                                                              & 9      & 58\\ \hline 

\begin{tabular}[c]{@{}l@{}}``bigram:\_api=regcreatekeyexw+arguments=software\path{\\}wanacrypt0r",\\
``enhanced:\_object=dir+event=create+data=file:\\ 
c:\path{\\}docume$\sim$1\path{\\}cuckoo\path{\\}locals$\sim$1\path{\\}temp\path{\\}msg",\\
``enhanced:\_object=file+event=execute+data=file:attrib +h . ",\\
``enhanced:\_object=file+event=execute+data=file:icacls . /grant everyone:f /t /c /q ",\\ ``enhanced:\_object=file+event=write+data=file:\\
c:\path{\\}docume$\sim$1\path{\\}cuckoo\path{\\}locals$\sim$1\path{\\}temp\path{\\}00000000.pky",\\
``enhanced:\_object=file+event=write+data=file:\\
c:\path{\\}docume$\sim$1\path{\\}cuckoo\path{\\}locals$\sim$1\path{\\}temp\path{\\}r.wnry" \end{tabular} & 16      & 80                      \\ \hline 

\end{tabular}
\end{lrbox}
\scalebox{0.65}{\usebox{\tablebox}}
\vspace{-.4cm}
\end{table}

\subsection{Analysis of Polymorphic WannaCry Malware}
As introduced in Section~\ref{sec:method}, Cuckoo Sandbox analyzes all the behavioral activities of the submitted files while searching on \url{VirusTotal.com} for matching 63 AV vendor's signatures with the suspicious file. 
The WannaCry malware is identified in Experiment~\ref{subsec:experiment1} and \ref{subsec:experiment2} since we submitted a single WannaCry executable file to Cuckoo Sandbox. 
In Experiment \ref{subsec:experiment3}, we combined normal activities and the malware executable file into one Python file. 
Although the same WannaCry executable is called by the Python script, 0 of the 63 AV vendors alert on it. 
We conjecture this is because the content of the Python file has no malware patterns. 
Although our method can still identify the malware in this case, we design a final experiment to validate that our TF-IDF method can identify more subtle polymorphism. 

To create a very similar variant, we modify the HEX code file of the WannaCry malware; specifically, we change the upper-case letters of the message \textit{``This program cannot be run in DOS mode.''} to all lower-case letters and the space characters to ``-". 
The polymorphic WannaCry executable file is then submitted to Cuckoo Sandbox, and none of the 63 virus databases of the VirusTotal scanner finds matched signatures of the polymorphic WannaCry malware. 
By using the same technique with the same four normal activity analysis reports as shown in Experiment One (Section~\ref{subsec:experiment1}), the features and weights calculated by using the polymorphic malware analysis report but are the same as Table~\ref{weight1}.



\section{Conclusion and Future Work}
\label{sec:conclusion}
In this paper, we present a method to automatically extract features of malware from host logs. 
Our experiments employed the relatively new and impactful WannaCry ransomware. 
For empirical validation we employed behavior logs from the analysis reports generated by Cuckoo Sandbox under various scenarios of normal and malware activities. 
Our experimental results validate that the method can extract distinguishing features of the malware from logs containing a majority of non-malicious events, and is robust to polymorphism.
Most importantly, given a majority of ambient logs with ransomware activities also included, accurate extraction of many ransomware features are automatically identified. 

Furthermore, we have identified and empirically exhibited exactly how false indicators of malware could arise from our method\textemdash by non-malware features occurring in the malware document, but relatively infrequently otherwise.  
In practice, for creating patterns from dynamic analysis, this scenario is easily avoided. 
Testing the malware analysis and pattern generation capability on ambient logs collected by operations will be next-step research. 
Our results in Experiment~\ref{subsec:experiment3} indicate that in these adverse scenarios, although some highly ranked features may be spurious, the majority of the $\approx$40 top-ten ranked features are accurate indicators. 

Although presentation of the method and results is outside the scope of this paper, the TF-IDF approach gives better results for analyzing WannaCry malware than other discriminant analysis algorithms based on Fisher's Linear Discriminant Analysis~\cite{welling2005fisher}. 
Further, the preservation of understandable features is a prerequisite for automating malware analysis that TF-IDF provides that other analysis capabilities  do not.

Future research will consider integration with other detection systems~\cite{harshaw2016graphprints, ferragut2012new} for automatic pattern generation,  or enhancing autonomic security systems~\cite{chen2014autonomic, Chen2013A}. 
Overall, we hope this contribution leads to operational implementations to expedite manual analysis of logs, malware analysis, and to provide accurate pattern generation  from both dynamic analysis tools and host logs. 



\bibliographystyle{ieeetr}
\bibliography{ref}

\break
\clearpage
\section{Appendix (Supplemental Information)} 
\addcontentsline{toc}{section}{Appendix}
\label{sec:appendix}  

\subsection{WannaCry Ransomware Details} 
\label{sec:appendix-wannacry}
The encryption component imports CryptoAPI from \texttt{advapi32.dll}, makes a file copy of itself, and extracts a zip archive from the encryptor's resource section~\cite{WannaCry2017}. 
The zip archive contains six \texttt{.wnry} files, a folder and two executable files, and they are
\begin{enumerate}
	\item \texttt{b.wnry}: bitmap file used as the victim's desktop wallpaper
	\item \texttt{c.wnry}: config  file with websites, target addresses, and Tor communication endpoints
	\item \texttt{s.wnry}: Tor client 
	\item \texttt{t.wrny}: \texttt{WANACRY!} file containing default public and private keys 
	\item \texttt{u.wnry} \texttt{@WannaDecryptor@.exe} file
	\item \texttt{r.wnry} Q\&A file with payment instructions 
	\item \texttt{\textbackslash msg} folder with 128 RTF files in different languages to inform victims their data is encrypted and give instructions to decrypt the files.
	\item \texttt{taskse.exe}: file for launching the decryption tool 
	\item  \texttt{taskdl.exe} executable file for removing temporary files with \texttt{.WNCRYT} extension in the current folder that has the executable file and the Recycle Bin folder
\end{enumerate}

The WannaCry malware then generates a unique identifier based on the name of the victim machine. The unique identifier consists of 8-15 random lowercase characters followed by three numbers. For example, the WannaCry malware unique identify generated for a Cuckoo Sandbox with a victim Windows XP computer used for our experiment is \textit{thsgvkvtwaipdcd971}.

\begin{figure}[!ht]
\includegraphics[scale=0.25]{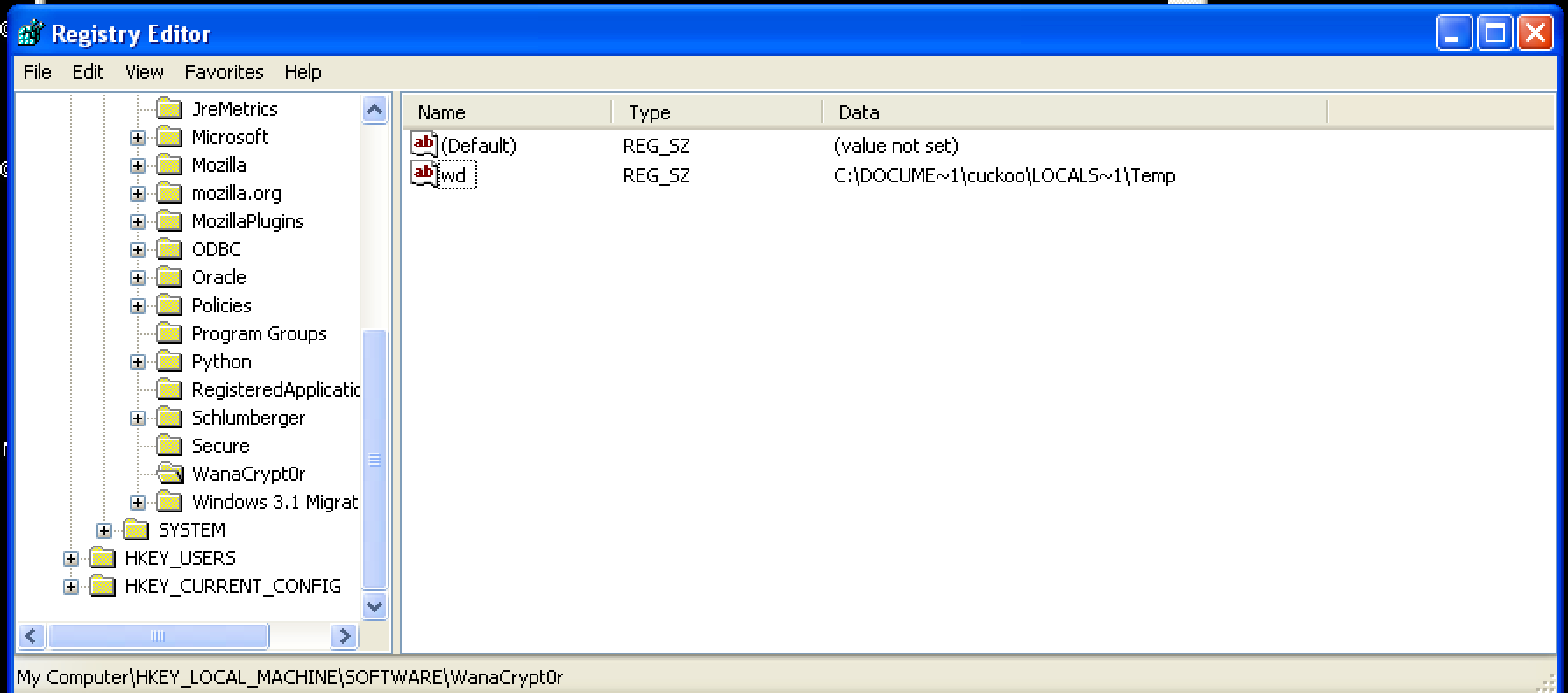}
\centering
\caption{Set Current Directory to a New Registry Key by the WannaCry Malware }
\label{fig:registrykey}
\end{figure}

The current directory where the WannaCry malware is located is updated by the malware and it also creates a registry \path{HKEY_LOCAL_MACHINE\Software\WanaCrypt0r\wd} and sets its value to the current directory (See Figure~\ref{fig:registrykey}), e.g., \path{C:\DOCUME~1\cuckoo\LOCALS~1\Temp}. 

\begin{figure}[!ht]
\includegraphics[scale=0.3]{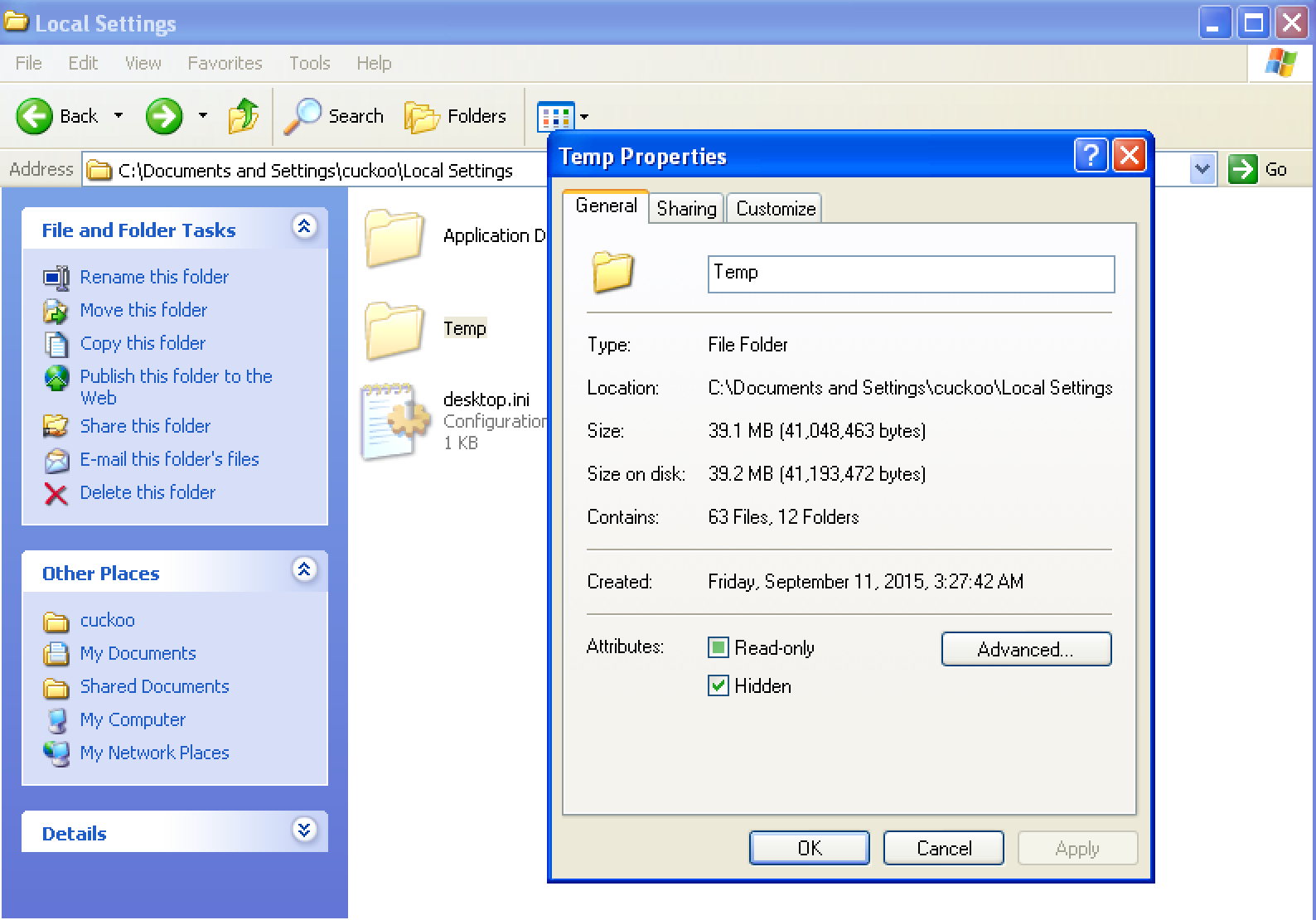}
\centering
\caption{Hide the Folder Containing Malware}
\label{fig:hiddenfolder}
\end{figure}

\begin{figure}[!ht]
\includegraphics[scale=0.3]{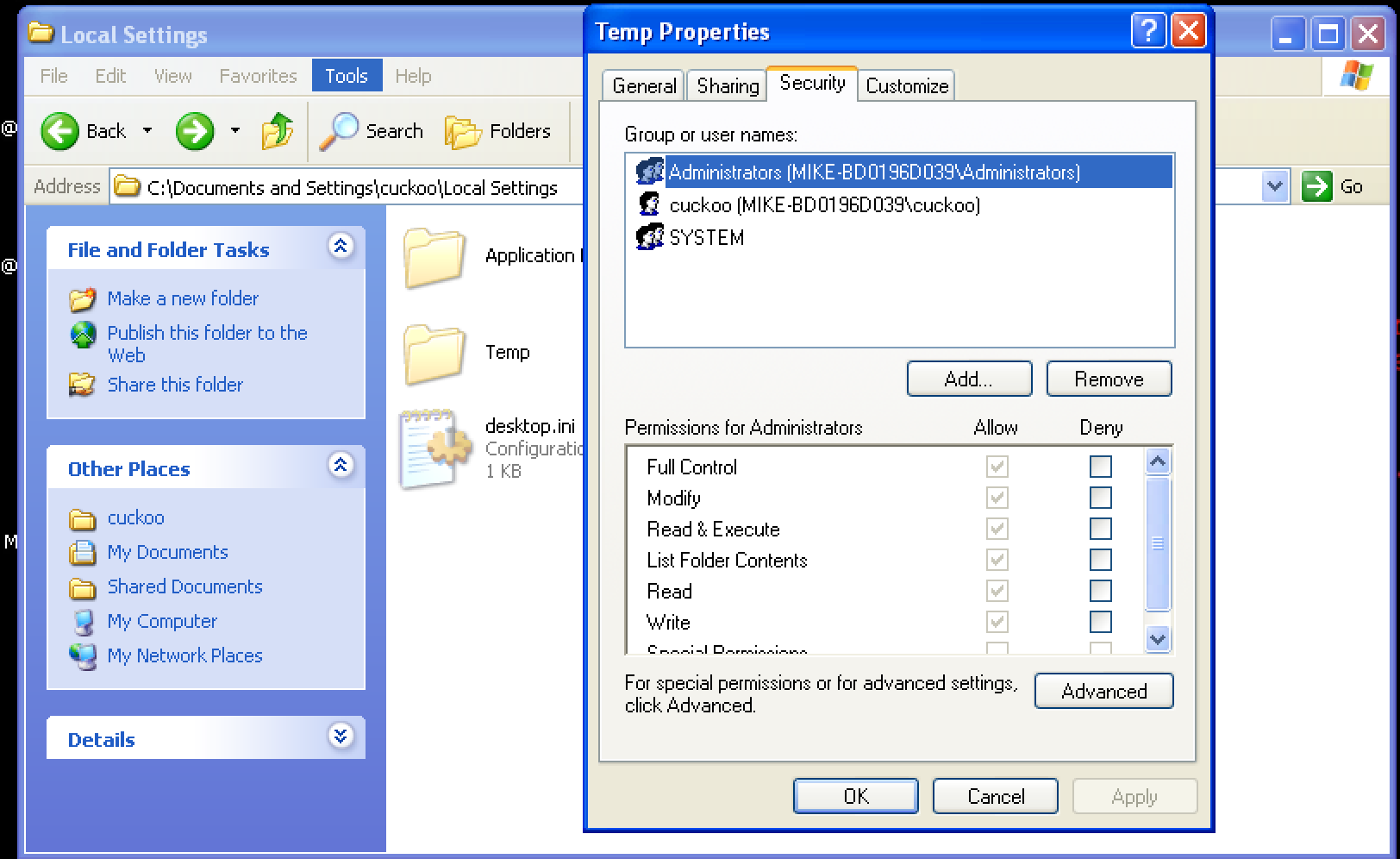}
\centering
\caption{Grant All User Permissions to the Malware Folder}
\label{fig:permission}
\end{figure}

The configuration file, \textit{c.wnry}, which contains websites, targeted addresses and Tor communication endpoints is modified by the malware. A string ``12t9YDPgwueZ9NyMgw519p7AA8isjr6SMw" is added to the original configuration file. After that, the malware sets the current directory \path{ C:\DOCUME~1\cuckoo\LOCALS~1\Temp } as a hidden folder by executing the \textit{attrib +h .} command  (see Figure~\ref{fig:hiddenfolder}. The current directory and their sub-directories are granted all user permissions by the malware with executing this command: ``icacls . /grant Everyone:F /T /C /Q" (check Figure~\ref{fig:permission})~\cite{CERT2017WannaCry}.

The malware then imports the RSA AES key from the t.wnry file, and loads a Win32 PE DLL into memory to start encrypting files by calling \textit{TaskStart}. The WannaCry malware generates public and private RSA keys (e.g.,  \texttt{00000000.pky} and \texttt{00000000.eky}), and saved them into the current directory. After that, the malware keeps writing 136 bytes including current time of the system to the file, \texttt{00000000.res}, every 25 seconds. 

The malware uses \texttt{SHGetFolderPathW} API to find target files in the hard drive (expect CDROM) and scans for new drives attached to the system every three seconds.  One AES key per target file is generated, and the public RSA key is used to encrypt AES keys. \texttt{CreateFileW}, \texttt{ReadFile} and \texttt{WriteFile} APIs are called to create encrypted files. The string \texttt{WANNACRY} is written on the infected files.

Note that files in the shared folder of the host are also encrypted. Meanwhile, the malware starts a thread to execute \texttt{taskdl.exe} every 30 seconds to replace the temporary encrypted files with an \texttt{.WNCRYT} extension to \texttt{.WCRY}.

The malware executes the command: 
``\texttt{taskse.exe} \path{C:\DOCUME~1\cuckoo\LOCALS~1\Temp\@WanaDecryptor@.exe}" to launch the decryption tool. A registry key named with the unique identify (i.e., thsgvkvtwaipdcd971) is created by the malware using the command \path{cmd.exe /c reg add HKLM\SOFTWARE\Microsoft\Windows\CurrentVersion\Run /v \"thsgvkvtwaipdcd971\" /t REG_SZ /d \"\"C:\DOCUME~1\cuckoo\LOCALS~1\Temp\tasksche.exe\\"\" /f}

The \texttt{@WanaDecryptor@.exe} is then executed, and the updates of the bitcoin address is saved in \texttt{c.wnry}. The file \texttt{u.wnry} is then copied to \texttt{@WanaDecryptor@.exe}, and the contents of \texttt{r.wnry} are copied to the \texttt{@WanaDecryptor@.txt} file.
The WannaCry malware scans the victim's desktop and documents folders. Temporary files starting with ``$\sim$SD" are created then deleted automatically. 

Once the malware completes the encryption process, it executes \texttt{taskkill.exe} to kill the processes of Microsoft Exchange and SQL. The malware also encrypts files on logical drives and replaces the desktop image to \texttt{!WannaCryptor!.bmp}, a copy of \texttt{b.wnry}.

The most significant features of the WannaCry malware summarized from the above technical analysis are shown in Table~\ref{tab:actions}. 
These features are separated into two categories, Pre-Encryption Features and Encryption Features. From the WannaCry technical analysis, we can find that the file systems will not be encrypted until the malware generates the private key \texttt{00000000.eky}. Therefore, the features extracted from activities which happen before the production of the private key are identified as Pre-Encryption Features, and all features after the application of the private key are considered as Encryption Features. These Pre-Encryption Features are  essential to analyze and detect the early behaviors of the WannaCry malware.

\subsection{Cuckoo Output Categories} 
\label{sec:appendix-cuckoo}
The JSON Cuckoo analysis report generated is saved into seven main categories, and they are:
\begin{enumerate}
	\item signatures: users are allowed to predefined patterns of known malware. If the analyzed malware matches the patterns, a new entry can be found in the ``signatures" category. The value of ``signatures" remains empty if the analyzed malware are unknown.

	\item virustotal: Cuckoo searches on VirusTotal.com for antivirus signatures of the analyzed file. 63 engines, such as McAfee and Kaspersky, are scanned for identifying the malware. 

	\item static: the static analysis module analyzes PE32 files and provides version information, sections, resources and libraries imported by the analyzed file.
	\item dropped: this category presents information of files that are dropped by the analyzed file and dumped by Cuckoo, including temporary files which are eventually deleted by the malware. 
	\item network: Cuckoo also monitors and records real-time network traffic into PCAP files during the analysis. Network information such as source and destination IP addresses and port numbers, DNS traffic, hosts, HTTP requests, IRC, SMTP traffic are extracted and saved into the JSON report file.
	\item behavior: the raw behavioral logs for each process running by the analyzed files are transformed and interpreted. This category includes logs of the complete processes tracing, a behavioral summary and a process tree. The \textbf{anomaly} class under the behavior module detects activities such as removing Cuckoo's hooks, and mark the unhook activities as anomalies.
The \textbf{enhanced} class generates a more extensive high-level summary of the processes and their activities. Instead of reading from raw behavioral logs, the enhanced class helps to interpret and summarize essential activities performed by the analyzed files. Information generated by the enhanced class includes activities such as read, write and delete registry keys, files and directories; load Windows libraries; and execute files. In our experiment, we extract features from the logs generated and interpreted by the enhanced class. We check the raw behavioral logs only if the extracted enhanced features are not sufficient to identify the patterns of the malware.

\item volatility: this category shows the memory dump analysis results.

\end{enumerate}
 
\end{document}